\documentstyle[prd,aps,floats,tabularx,epsfig]{revtex}
\newcommand{\be}{\begin{equation}}
\newcommand{\ee}{\end{equation}}
\newcommand{\bea}{\begin{eqnarray}}
\newcommand{\eea}{\end{eqnarray}}

\begin{document}
\setlength{\unitlength}{1mm}
\twocolumn[\hsize\textwidth\columnwidth\hsize\csname@twocolumnfalse\endcsname
\title{Current constraints on Cosmological Parameters from Microwave
Background Anisotropies.}
\author{Alessandro Melchiorri$^\flat$ and Carolina J. \"Odman$^\sharp$}
\address{ 
$^\flat$ Astrophysics, Denys Wilkinson Building, University of Oxford, Keble ro
ad, OX1 3RH, Oxford, UK\\
$^\sharp$ Astrophysics Group, Cavendish Laboratory, Cambridge University, Cambridge, U.K.\\}
\maketitle

\begin{abstract}
We compare the latest observations of Cosmic Microwave Background 
(CMB) Anisotropies
with the theoretical predictions of the standard scenario
of structure formation.
Assuming a primordial power spectrum of adiabatic perturbations
we found that the total energy density is constrained to be 
$\Omega_{tot}=1.03\pm0.06$ while the energy density in baryon
and Cold Dark Matter (CDM) are  $\Omega_bh^2=0.021\pm0.003$ and
$\Omega_{cdm}h^2=0.12\pm0.02$, (all at $68 \%$ C.L.) respectively.  
The primordial spectrum is consistent with scale invariance, 
($n_s=0.97\pm0.04$) and the age of the universe is $t_0=14.6\pm0.9$ Gyrs.
Adding informations from Large Scale Structure and Supernovae, we found
a strong evidence for a cosmological constant 
$\Omega_{\Lambda}=0.70_{-0.05}^{+0.07}$ and a value of the Hubble
parameter  $h=0.69\pm0.07$.
Restricting this combined analysis to flat universes, we put 
constraints on possible 'extensions' of the standard scenario.
A gravity waves contribution to the quadrupole anisotropy is limited
to be $r \le 0.42$ ($95 \%$ c.l.). 
A constant equation of state for the dark energy
component is bound to be $w_Q \le -0.74$ ($95 \%$ c.l.). 
We constrain the effective relativistic degrees of freedom 
$N_\nu \leq 6.2$ and the neutrino chemical potential 
$-0.01 \leq \xi_e \leq 0.18$ and $|\xi_{\mu,\tau}|\leq 2.3$
(massless neutrinos). 
\end{abstract}

\bigskip]

\section{Introduction}

The last years have been an exciting period for
the field of the CMB research.
With recent CMB balloon-borne and ground-based experiments we are 
entering a new era of 'precision' cosmology that enables us to use 
the CMB anisotropy measurements to constrain the cosmological parameters 
and the underlying theoretical models.
With the TOCO$-97/98$ (\cite{torbet},\cite{miller}) 
and BOOMERanG-$97$ (\cite{mauskopf}) experiments a firm detection of
a first peak on about degree scales has been obtained. 
In the framework of adiabatic Cold Dark Matter (CDM) models, the
position, amplitude and width of this peak provide strong supporting 
evidence for the inflationary predictions of
a low curvature (flat) universe and a scale-invariant primordial 
spectrum (\cite{knox}, \cite{melchiorri}, \cite{tegb97}).

The new experimental data from BOOMERanG LDB (\cite{netterfield}), 
DASI (\cite{halverson}), MAXIMA (\cite{lee}), CBI (\cite{pearson}), 
VSA (\cite{scott}) and, more recently, ACBAR (\cite{acbar}), 
ARCHEOPS (\cite{benoit}) and revised and improved analysis
from BOOMERanG (\cite{ruhl}) and VSA (\cite{vsae})
have provided further evidence for the presence of 
the first peak and refined the data at larger multipole
(see e.g. \cite{carolina}). 
The combined data suggest the presence of 
a second and third peak in the spectrum, confirming the model
prediction of acoustic oscillations in the primeval plasma 
and sheding new light on various cosmological and 
inflationary parameters (\cite{debe01}, \cite{wang}, \cite{pryke})
\footnote{However, it is important to notice that datasets
appeared before April 2001 do not show presence of multiple
peaks (see \cite{podariu}) and that analsyses of the most recent
 datasets not based on Bayesian methods can give weaker constraints
on the peak amplitude and positions (see e.g. \cite{cmiller})}.

In this {\it Rapid Communication} we compare the 
latest measurements of the Cosmic Microwave Background 
Anisotropies angular power 
spectrum with the theoretical predictions of the standard
CDM scenario in order to constrain most of its parameters.

Similar and careful analysis have been done recently
(\cite{lewisbridle}, \cite{pearson}, \cite{melksilk}), 
the work presented here can be considered as a last-minute 
update of most of the results 
already published but it will also differ for following aspects:
First of all, we will include the new ARCHEOPS, ACBAR, BOOMERanG and 
VSAE datasets, which provide the best determination to the date of 
region in the spectrum from the scales sample to COBE up to the 
Silk damping scales.

Second, we will also focus on possible deviations to the standard
scenario, like gravity waves, an equation
of state for the dark energy $w_Q > -1$ or an extra-background
of relativistic particles.

Our paper is then organized as follows: In section II 
we present the analysis method we used. In Section III
we report our results. In Section IV we discuss our
conclusions.

\section{Method}

\begin{figure}
\begin{center}
\includegraphics[scale=0.40]{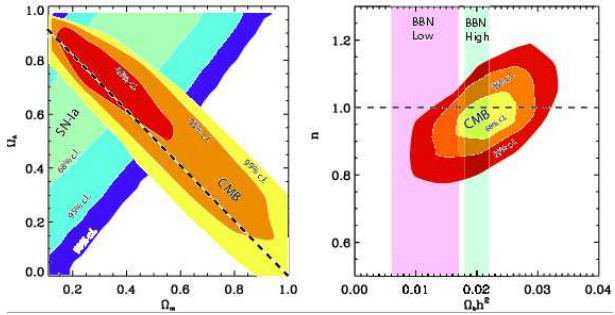}
\end{center}
\caption{Confidence contours in the $\Omega_M - \Omega_{\Lambda}$ 
(left) and $\Omega_bh^2 - n$ (Right) planes from the analysis described 
in the text.}
\label{fig3}
\end{figure} 

As a first step, we consider a template of adiabatic, 
$\Lambda$-CDM models computed with CMBFAST (\cite{sz}), 
sampling the various parameters as follows:
the physical density in cold dark matter 
$\Omega_{cdm}h^2\equiv \omega_{cdm}= 0.01,...0.40$, in steps of  $0.01$;  
the physical density in baryons 
$\Omega_{b}h^2\equiv\omega_{b} = 0.001, ...,0.040$, 
in steps of  $0.001$, the cosmological constant 
$\Omega_{\Lambda}=0.0, ..., 0.95$, 
in steps of  $0.05$ and the curvature 
$\Omega_{k}=-0.5,....0.5$ step $0.05$.
The value of the Hubble constant is not an independent 
parameter, since: 
 \begin{equation} 
h=\sqrt{{\omega_{cdm}+\omega_b} \over {1-\Omega_{\Lambda}-\Omega_k}}. 
\end{equation}
We allow for a reionization of the intergalactic medium by
varying also the compton optical depth parameter 
$\tau_c$ in the range $\tau_c=0.0,...,0.45$ in steps of $0.05$
\footnote{We point out that values of $\tau_c > 0.20$ are in disagreement 
with recent estimates of the redshift 
of reionization $z_{re}\sim 6 \pm 1$ (see e.g. \cite{gnedin})
which point towards $\tau_c \sim 0.05$.
However, since the reionization mechanisms is still unclear,
we prefer to consider also greater values of $\tau_c$.}.

We also vary the scalar spectral index of primordial fluctuations
in the range $n_S=0.7,...,1.3$ in steps of $0.02$.

We will then restrict our analysis to {\it flat} models and,
adding external priors as described below, we will constrain
possible extensions of the standard model.
In particular, we will consider a background
of gravity waves, parametrized as a contribution to the
CMB anisotropy quadrupole $r=C_2^T/C_2^S$. We consider the 
tensor spectral index $n_T$ to be $n_T=-r/6.8$ for $n_S < 1$ 
and $n_T=0$ for $n_S > 1$.

We will also consider an equation of state for the dark energy $w_Q \neq -1$
sampled as $w_Q=-1.0,...,-0.4$ in step of $0.05$.
Finally we will constrain an extra-background of relativistic
particles, parametrized through an effective number of
relativistic neutrinos $\Delta N^{eff}$ sampled as
$\Delta N^{eff}=0.0,...,10.0$ in step of $0.5$.

For the CMB data, we use the recent results from the 
BOOMERanG-98, DASI, MAXIMA-1, CBI, VSAE, ACBAR and ARCHEOPS experiments. 
Where possible, we use the publicly available window functions 
and offset lognormal correction prefactors $x_b$ in order 
to compute the theoretical band power signal $C_B$ as in \cite{BJK}.
The likelihood for a given theoretical model is defined by 
 $-2{\rm ln} {\cal L}=(C_B^{th}-C_B^{ex})M_{BB'}(C_{B'}^{th}-C_{B'}^{ex})$ 
where  $M_{BB'}$ is the Gaussian curvature of the likelihood  
matrix at the peak.

We include the beam and calibration uncertainties by the marginalization 
methods presented in (\cite{bridle}, see also \cite{ganga}).

In addition to the CMB data we will also consider 
the real-space power spectrum  
of galaxies in the 2dF 100k galaxy redshift survey using the 
data and window functions of the analysis of Tegmark et al. (\cite{thx}). 
 
To compute ${\cal L}^{2dF}$, we evaluate $p_i = P(k_i)$,  
where $P(k)$ is the theoretical matter power spectrum  
and $k_i$ are the $49$ k-values of the measurements in \cite{thx}.  
Therefore we have $-2ln{\cal L}^{2dF} = \sum_i [P_i-(Wp)_i]^2/dP_i^2$, 
where $P_i$ and $dP_i$ are the measurements and corresponding error bars 
and $W$ is the reported $27 \times 49$ window matrix. 
We restrict the analysis to a range of scales where the fluctuations 
are assumed to be in the linear regime ($k < 0.2 h^{-1}\rm Mpc$). 
When combining with the CMB data, we marginalize over a bias $b$  
considered to be an additional free parameter. 

Furthermore, we will also incorporate constraints obtained
from the luminosity measurements of type I-a supernovae (SN-Ia).
The observed apparent bolometric luminosity 
is related to the luminosity distance, measured in Mpc, by
$m_{B}=M+5 log d_{L}(z)+25$.
where M is the absolute bolometric magnitude. 
The luminosity distance is sensitive to the cosmological 
evolution through an integral dependence on the Hubble factor 
$d_{l}=(1+z)\int_{0}^{z} (dz'/H(z',\Omega_{Q},\Omega_{M},w_{q})$ 
where $\Omega_{Q}$ and $w_{Q}$ are the energy density
and equation of state of the ``dark energy'' component. 
We evaluate the likelihoods 
assuming a constant equation of state, such that 
$H(z)=\rho_{0}\sum_{i}\Omega_{i}(1+z)^{(3+3w_{i})}$. 
The predicted $m_{eff}$ is then calculated by calibration 
with low-z supernovae observations  where 
the Hubble relation $d_{l}\approx H_{0}cz$ is obeyed. 
We calculate the likelihood, ${\cal L}$, using the relation 
${\cal L}={\cal L}_{0}\exp(-\chi^{2}(\Omega_Q,\Omega_M,w_Q)/2)$ 
where ${\cal L}_{0}$  is an arbitrary normalisation 
and $\chi^{2}$ is evaluated using the observations of (\cite{super1}),
marginalising over $H_{0}$.

Finally, we will also consider a $1-\sigma$ contraint on the Hubble 
parameter, $h=0.71\pm0.07$, derived from Hubble Space Telescope 
(HST) measurements (\cite{freedman}).

In order to constrain a parameter $x$ we marginalize over 
the values of the other parameters $\vec{y}$. This yields the marginalized
likelihood distribution

\begin{equation}
{\cal L}(x) {\equiv}
        P({x}| \vec{\cal C}_B) =
        \int {\cal L}(x,\vec{y})  d \vec{y}.
\end{equation}

\noindent The central values and $1\sigma$ limits are then found from 
the 16\%, 50\% and 84\% integrals of $\mathcal{L}(x)$.

\section{Results: Standard Parameters}

In Fig. $1$ we plot the likelihood contours on the 
$\Omega_{M}-\Omega_{\Lambda}$ and $\Omega_bh^2-n_S$ planes, 
using only the CMB data.

As we can see from Figure $1$ (Left Panel)
the data strongly suggest a flat universe
(i.e. $\Omega=\Omega_M+\Omega_{\Lambda}=1$). From our CMB dataset 
we obtain $\Omega=1.03\pm0.06$ at $95 \%$ C.L..

The inclusion of complementary datasets in the analysis
breaks the angular diameter distance degeneracy 
and provides evidence for a cosmological constant at high significance.
Adding the HST constraint, the 2dF dataset 
and SN-Ia gives $\Omega_{\Lambda}=0.67_{-0.13}^{+0.07}$, 
$\Omega_{\Lambda}=0.63_{-0.09}^{+0.11}$ and 
$\Omega_{\Lambda}=0.70_{-0.05}^{+0.07}$, all at $68 \%$ c.l..

Combining CMB and 2dF gives $h=0.69 \pm 0.07$ in extremely
good agreement with the HST result.

In the right panel of Fig.1 we plot the CMB likelihood contours in the 
$\Omega_bh^2-n_S$ plane. 
As we can see, the present CMB data is in beautiful agreement with 
{\it both} a nearly scale invariant
spectrum of primordial fluctuations, as predicted by inflation, and
the value for the baryon density $\omega_b =0.020\pm0.002$ 
($95 \%$ C.L.)  predicted
by Standard Big Bang Nucleosynthesis (see e.g. \cite{burles}) from
measurements of primordial deuterium.
For the scalar spectral index, we found: $n_s=0.97\pm0.04$.
However, the CMB constraint is also in agreement in between $2 - \sigma$ 
with the lower
BBN value $0.006< \Omega_bh^2 < 0.017$ obtained from measurements
of $^4He$ and $^7Li$ (\cite{cyburt}) at $95 \%$ C.lL..

An increase in the optical depth $\tau_c$ after recombination 
by reionization (see e.g. \cite{haiman} for a review) or by some more
exotic mechanism damps the amplitude of the CMB peaks.
Degeneracies with other parameters such as $n_S$ are present
(see e.g. \cite{debe97}) and we cannot strongly bound the value 
of $\tau_c$. In the range of parameters we considered we 
have $\tau_c \le 0.24$ at $1-\sigma$.

The amount of non-baryonic dark matter is also 
constrained by the CMB data with $\Omega_{dm}h^2=0.12 \pm 0.02$ 
at $68 \%$ c.l..
The presence of power around the third peak is crucial in this sense,
since it cannot be easily accommodated in models based on just baryonic
matter (see e.g. \cite{melksilk}, \cite{lmg}, \cite{mcgaugh} 
and references therein).

Furthermore, under the assumption of flatness, we can derive important
constraints on the age of the universe $t_0$.
From our cmb dataset we obtain $t_0=14.6 \pm 0.9$ GYrs 
consistent with the analyses of (\cite{iggy}, \cite{netterfield}, 
\cite{knoxage}).

\section{Results: Beyond the Standard Model.}

As discussed before, even if the present CMB observations 
can be fitted with just $5$ parameters it is interesting to extend 
the analysis to other parameters allowed by the theory.
Here we will just consider a few of them.

{\bf Gravity Waves.}
 
The metric perturbations created during inflation belong to two types:
{\it scalar} perturbations, which couple to the stress-energy of 
matter in the universe and form the ``seeds'' for structure formation 
and {\it tensor} perturbations, also known as 
gravitational wave perturbations.
Both scalar and tensor perturbations contribute to CMB anisotropy.
In most of the recent CMB analysis the tensor modes have been neglected, 
even though a sizable background of gravity waves 
is expected in most of the inflationary scenarios. 
Furthermore, in the simplest models,
a detection of the GW background 
can provide information on the second derivative
of the inflaton potential and shed light on the physics at
$\sim 10^{16} Gev$ (see e.g. \cite{hoffman}).

The shape of the $C^T_{\ell}$ spectrum from tensor modes is drastically
different from the one expected from scalar fluctuations,
affecting only large angular scales (see e.g. \cite{crittenden}). 
The effect of including tensor modes is similar to 
just a rescaling of the degree-scale $COBE$ normalization and/or 
a removal of the corresponding data points from the analysis.

This further increases the degeneracies among cosmological
parameters, affecting mainly the estimates of the baryon and 
cold dark matter densities and the scalar spectral index $n_S$
(\cite{melk99},\cite{kmr}, \cite{wang}, \cite{efstathiougw}).

The amplitude of the GW background is therefore weakly constrained
by the CMB data alone, however, when information from galaxy clustering 
and SN-Ia are included, an upper limit on $r$ can be obtained. 

In Fig.2 we plot the contraints obtained in the $n_S-r$ plane
under the assumption of flatness and including the 2dF and SN-Ia data.
As we can see, the possibility of a tensor component is still
in agreement with the combined analysis of these datasets.
Including a conservative BBN constraint $\omega_b <0.024$ 
further improves the bound to $r \le 0.42$ at $95 \%$ c.l..
Similar bounds have been found in a previous analysis (\cite{carolina2}),
without the VSAE, ACBAR and Boomerang revised datasets.

\begin{figure}
\begin{center}
\includegraphics[scale=0.35]{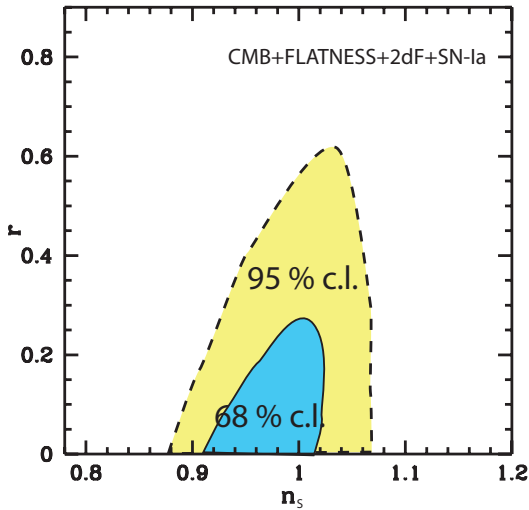}
\end{center}
\caption{The $68 \%$ and $95\%$ CL contours for the gravity waves
contribution in the $n_S-r=C_2^T/C_2^S$ plane.}
\label{gravity}
\end{figure}

{\bf Quintessence.}

The discovery that the universe's evolution may be dominated by
an effective cosmological constant \cite{super1}
is one of the most remarkable cosmological findings of recent years.
One candidate that could possibly explain the observations is a
dynamical scalar ``quintessence'' field. 
The common characteristic of quintessence
models is that their equations of state, $w_{Q}=p/\rho$, vary with time
while a cosmological constant remains fixed at
$w_{Q=\Lambda}=-1$. 
Observationally distinguishing a time variation in
the equation of state or finding $w_Q$ different from $-1$ will
therefore be a success for the quintessential scenario.
Quintessence can also affect the CMB by acting as an additional 
energy component with a characteristic viscosity.
However any early-universe imprint of quintessence 
is strongly constrained by Big Bang Nucleosynthesis
with $\Omega_Q(MeV) < 0.045$ at $2\sigma$ for temperatures near 
$T \sim 1Mev$ (\cite{bhm}).

\begin{figure}
\begin{center}
\includegraphics[scale=0.35]{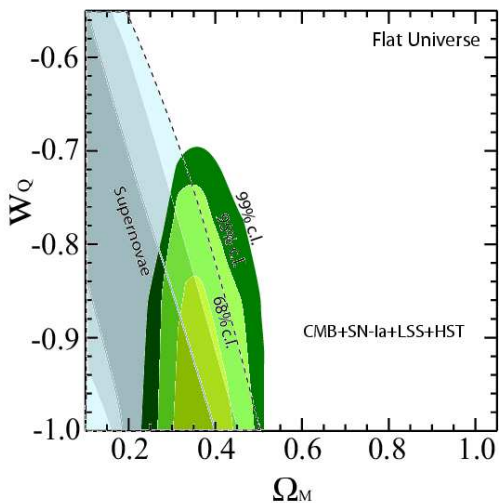}
\end{center}
\caption{The likelihood contours in the ($\Omega_M$, $w_Q$) plane,
with the remaining parameters taking their best-fitting values for the
joint CMB+SN-Ia+2dF+HST analysis described in the text.
The contours correspond to the 68\%, 95\% and 99\% confidence levels 
respectively.}
\label{figo1}
\end{figure}

In Figure 3 we plot the likelihood contours in the ($\Omega_M$, $w_Q$) plane 
from our joint analyses of CMB+SN-Ia+HST+2dF together with the contours 
from the SN-Ia dataset only.
The new CMB results improve the constraints from previous and 
similar analysis (see e.g., \cite{PTW}), 
\cite{rachel}, \cite{mortsell}) with $w_Q<-0.87$ at $68 \%$ c.l..
The current constraints are then perfectly in agreement with the $w_Q=-1$
cosmological constant case and gives no support to a
quintessential field scenario with $w_Q > -1$.

In our analysis we only consider the case of a constant-with-redshift
$w_Q$. The assumption of a constant $w_Q$ is based on
several considerations: first of all, since
both the luminosities and angular distances (that are
the fundamental cosmological observables)
depend on $w_Q$ through multiple
integrals, they are not particularly sensitive to
variations of $w_Q$ with redshift. Therefore, with current data, 
no strong constraints can be placed on the redshift-dependence of $w_Q$.
Second, for most of the dynamical models on the market,
the assumption of a piecewise-constant equation of state is
a good approximation for an unbiased determination
of the effective equation of state
\begin{equation}
w_{\rm eff} \sim \frac{\int w_Q(a) \Omega_Q(a) da}{\int \Omega_Q(a) da}
\end{equation}
predicted by the model.
Hence, if the present data is compatible with a constant
$w_Q=-1$, it may not be possible to discriminate between a cosmological
constant
and a dynamical dark energy model.

However one should be be very careful about drawing definitive 
conclusions about dark energy, since a constant
equation of state is still an approximation of a real
model of dark energy (see e.g. \cite{peeblesratra}). 
The analysis presented here should
be therefore regarded as a 'test' for deviations from 
the cosmological constant scenario.

{\bf Big Bang Nucleosynthesis and Neutrinos.}

As we saw in the previous section, the SBBN $95 \%$ CL region,
corresponding to $\Omega_b h^2= 0.020 {\pm} 0.002$ ($95 \%$ c.l.) (High BBN) 
and $0.006 < \Omega_b h^2 <0.017$ (Low BBN), 
 have a large overlap with the analogous CMBR contour. 
This fact, if it will be confirmed by future experiments on CMB
anisotropies, can be seen as one of the greatest
success, up to now, of the standard hot big bang model.

SBBN is well known to provide strong bounds on the number 
of relativistic species $N_\nu$. On the other hand,
Degenerate BBN (DBBN), first analyzed in Ref. \cite{d1,d2,Kang}, gives
very weak constraint on the effective number of massless neutrinos, since
an increase in $N_\nu$ can be compensated by a change in both the chemical
potential of the electron neutrino, $\mu_{\nu_e}= \xi_e T$,
and $\Omega_bh^2$. 
Practically, SBBN relies on the theoretical assumption that 
background neutrinos have negligible chemical potential, just like their 
charged lepton partners. Even
though this hypothesis is perfectly justified by Occam razor, models have
been proposed in the literature
\cite{AF,DK,DolgovRep,McDonald}, where large neutrino
chemical potentials can be generated. 

Combining the DBBN scenario with the bound on baryonic and radiation densities
allowed by CMBR data, it is possible to obtain strong constraints
on the parameters of the model. 
Such an analysis was, for example,  performed in (\cite{th7}, \cite{peloso}, 
\cite{hannestad}, \cite{orito}) 
using the first data release of BOOMERanG and MAXIMA 
(\cite{debe00}, \cite{hanany}). 

We recall that the neutrino chemical potentials contribute
to the total neutrino effective degrees of freedom $N_\nu$ as
\begin{equation}
N_{{\nu}} = 3 + \Sigma_{\alpha} \left[ \frac{30}{7}
\left( \frac{\xi_\alpha}{\pi} \right)^2 +
\frac{15}{7} \left( \frac{\xi_\alpha}{\pi} \right)^4 \right] \, .
\end{equation}
Notice that in order to get a bound on $\xi_\alpha$ we have here assumed   
that all relativistic degrees of freedom, other than photons, are given by
three (possibly) degenerate active neutrinos.

\begin{figure}
\begin{center}
\includegraphics[scale=0.45]{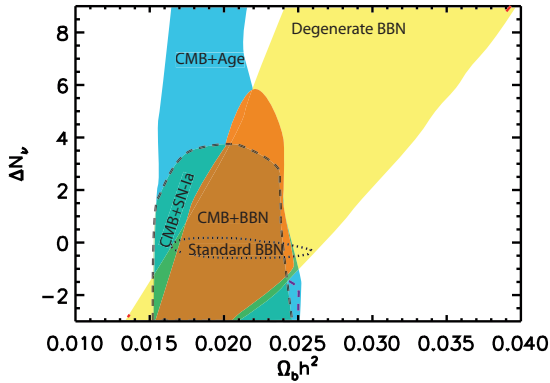}
\end{center}
\caption{The $95\%$ CL contours for degenerate BBN,
the CMB results with just the age prior, combined with SN-Ia and 
combined with BBN degenerate.}
\label{figDbbn}
\end{figure}

Figure~4 summarizes the main results with the new CMB data 
for the DBBN scenario (see caption). 
We plot the $95\%$ CL contours allowed by DBBN
together with the analogous $95 \%$ CL 
region coming from the CMB data analysis,
with only weak age prior, $t_0 > 11 $gyr and the
 $95 \%$ CL region of the joint product distribution ${\cal L} \equiv
{\cal L}_{DBBN}$${\cdot}{\cal L}_{CMB}$.

We obtain the bound $N_\nu \leq 8.4$, at $95 \%$ CL,
which translates into the bounds $-0.01\leq \xi_e \leq 0.24$, 
sensibly more stringent than what can be found from DBBN alone.
Combining CMBR and DBBN data with
the Supernova Ia data \cite{super1} strongly reduces the degeneracy
between $\Omega_m$ and $\Omega_\Lambda$. At $95 \%$ C.L. we 
find $\Delta N_\nu < 6.2 $, corresponding to  $-0.01\leq \xi_e \leq 0.18$ and
$|\xi_{\mu,\tau}|\leq 2.3$.  

It is however important to note that possible extra relativistic degrees 
of freedom, like light sterile neutrinos, would contribute to 
$N_{\nu}$ as well, and in this respect BBN cannot distinguish between 
their contribution to the total universe expansion
rate and the one due to neutrino degeneracy. Therefore, in a more general
framework, our estimates for $N_{\nu}$ can only represent an upper bound 
for the total neutrino chemical potentials.

Similar results have been obtained in \cite{hmmmp}, 
\cite{kneller} and \cite{hannestad2}.

\section{Conclusions}

The recent CMB data represent a beautiful success for the 
standard cosmological model. 
Furthermore, when constraints on cosmological parameters are 
derived under the assumption of adiabatic primordial perturbations
their values are in agreement with the predictions of the theory
and/or with independent observations.

As we saw in the previous section modifications 
as gravity waves, quintessence or extra background of relativistic
particles are still
compatible with current CMB observations, but are not necessary and
can be reasonably constrained when complementary datasets are included.

Since the inflationary scenario is in agreement with the data and all 
the most relevant parameters are starting to be constrained within a 
few percent accuracy, the CMB is becoming a wonderful laboratory for 
investigating the possibilities of new physics. With the promise of 
large data sets from Map, Planck and SNAP satellites and from the
SLOAN digital sky survey, opportunities may be open, for example, 
to constrain dark energy models, variations in fundamental constants 
and neutrino physics.

{\bf Acknowledgements}

We wish to thank Rachel Bean, Ruth Durrer, 
Steen Hansen, Pedro Ferreira, Mike Hobson, Will Kinney, Anthony Lasenby, 
Gianpiero Mangano, Gennaro Miele, Ofelia Pisanti, Antonio Riotto, 
Graca Rocha, Joe Silk, and Roberto Trotta for comments, 
discussions and help.


\begin{thebibliography}{}

\bibitem{AF} I. Affleck and M. Dine, {\it Nucl. Phys.} {\bf B249} (1985) 361.

\bibitem{bhm} R.~Bean, S.~H.~Hansen and A.~Melchiorri,
Phys.\ Rev.\ D {\bf 64} (2001) 103508
[arXiv:astro-ph/0104162].

\bibitem{rachel} R.~Bean and A.~Melchiorri,
arXiv:astro-ph/0110472, Phys.\ Rev.\ D Rapid Communication, in press.

\bibitem{benoit} A. Benoit {\it et al.} [Acheops Collaboration],
A \& A, submitted, 2002.

\bibitem{debe97} P.~de Bernardis, A.~Balbi, G.~De Gasperis, 
A.~Melchiorri and N.~Vittorio,
arXiv:astro-ph/9609154.

\bibitem{debe00} P.~de Bernardis {\it et al.}  [Boomerang Collaboration],
Nature {\bf 404}, 955 (2000)
[arXiv:astro-ph/0004404].

\bibitem{debe01} P.~de Bernardis {\it et al.}, [Boomerang Collaboration],
arXiv:astro-ph/0105296.


\bibitem{acbar}C.~l.~Kuo {\it et al.},
arXiv:astro-ph/0212289.

\bibitem{vsae} K.~Grainge {\it et al.},
arXiv:astro-ph/0212495.

\bibitem{ruhl} J.~E.~Ruhl {\it et al.},
arXiv:astro-ph/0212229.


\bibitem{bridle} S.~L.~Bridle, R.~Crittenden, A.~Melchiorri, M.~P.~Hobson, 
R.~Kneissl and A.~N.~Lasenby, arXiv:astro-ph/0112114.

\bibitem{BJK} J.~R.~Bond, A.~H.~Jaffe and L.~E.~Knox,
Astrophys.\ J.\  {\bf 533} (2000) 19
[arXiv:astro-ph/9808264].


\bibitem{burles}
S.~Burles, K.~M.~Nollett and M.~S.~Turner,
Astrophys.\ J.\  {\bf 552}, L1 (2001)
[arXiv:astro-ph/0010171].

\bibitem{cyburt}
R.~H.~Cyburt, B.~D.~Fields and K.~A.~Olive,
New Astron.\  {\bf 6} (1996) 215
[arXiv:astro-ph/0102179].

\bibitem{crittenden} R.~Crittenden, J.~R.~Bond, R.~L.~Davis, G.~Efstathiou and P.~J.~Steinhardt, 
Phys.\ Rev.\ Lett.\  {\bf 71} (1993) 324[arXiv:astro-ph/9303014].

\bibitem{DK} 
A.D. Dolgov and D.P. Kirilova, {\it J. Moscow Phys. Soc.} {\bf 1} (1991) 
217.                                                                          

\bibitem{DolgovRep}
A.D. Dolgov, {\it Phys. Rep.}{\bf 222} (1992) 309.

\bibitem{d1} A.G. Doroshkevich, I.D. Novikov, R.A. Sunaiev, Y.B. Zeldovich,
in {\it Highlights of Astronomy}, de Jager ed., (1971) p. 318.

\bibitem{knox} S.~Dodelson and L.~Knox,
Phys.\ Rev.\ Lett.\  {\bf 84}, 3523 (2000)
[arXiv:astro-ph/9909454].

\bibitem{efstathiougw} G. Efstathiou, astro-ph/0109151.

\bibitem{efsbond} G. Efstathiou \& J.R. Bond [astro-ph/9807103].

\bibitem{th7} S.~Esposito, G.~Mangano, A.~Melchiorri, G.~Miele and O.~Pisanti,
Phys.\ Rev.\ D {\bf 63} (2001) 043004
[arXiv:astro-ph/0007419].

\bibitem{iggy} I.~Ferreras, A.~Melchiorri and J.~Silk, 
MNRAS 327, L47 (2001), arXiv:astro-ph/0105384.

\bibitem{d2} W.A. Fowler,
Accademia Nazionale dei Lincei, Roma {\bf 157} (1971) 115.

\bibitem{freedman} W. Freedman {\it et al.}, 
Astrophysical Journal, 553, 2001, 47.

\bibitem{super1}  P.M. Garnavich et al, Ap.J. Letters \textbf{493}, L53-57
(1998); S. Perlmutter et al, Ap. J. \textbf{483}, 565 (1997); S.
Perlmutter et al (The Supernova Cosmology Project), Nature \textbf{391} 51
(1998); A.G. Riess et al, Ap. J. \textbf{116}, 1009 (1998);

\bibitem{gnedin}N. Gnedin, astro-ph/0110290.

\bibitem{lmg} L.~M.~Griffiths, A.~Melchiorri and J.~Silk,
Astrophys.\ J.\  {\bf 553} (2001) L5
[arXiv:astro-ph/0101413].

\bibitem{haiman} Z.~Haiman and L.~Knox,
arXiv:astro-ph/9902311.

\bibitem{halverson} N.~W.~Halverson {\it et al.},
arXiv:astro-ph/0104489.

\bibitem{hanany} S.~Hanany {\it et al.},
Astrophys.\ J.\  {\bf 545}, L5 (2000)
[arXiv:astro-ph/0005123].

\bibitem{hannestad}
S.~Hannestad, Phys.\ Rev.\ Lett.\  {\bf 85} (2000) 4203
[arXiv:astro-ph/0005018].

\bibitem{hannestad2} S.~Hannestad,
Phys.\ Rev.\ D {\bf 64} (2001) 083002
[arXiv:astro-ph/0105220].

\bibitem{Hansen} S.H. Hansen and F.L. Villante, {\it Phys. Lett.}
{\bf B486} (2000) 1.

\bibitem{hmmmp}
S.~H.~Hansen, G.~Mangano, A.~Melchiorri, G.~Miele and O.~Pisanti,
Phys.\ Rev.\ D {\bf 65} (2002) 023511
[arXiv:astro-ph/0105385].

\bibitem{hoffman} M. B. Hoffman, M. S. Turner, Phys.Rev. D64 (2001) 023506,
astro-ph/0006312.

\bibitem{Kang} H. Kang and G. Steigman,
{\it Nucl. Phys.} {\bf B372} (1992) 494.                                       

\bibitem{kmr}W.~H.~Kinney, A.~Melchiorri and A.~Riotto,
Phys.\ Rev.\ D {\bf 63} (2001) 023505[arXiv:astro-ph/0007375].

\bibitem{kneller} J.~P.~Kneller, R.~J.~Scherrer, G.~Steigman and T.~P.~Walker,
Phys.\ Rev.\ D {\bf 64} (2001) 123506
[arXiv:astro-ph/0101386].

\bibitem{knoxage} L. Knox, N. Christensen, C. Skordis,
[arXiv:astro-ph/0109232].

\bibitem{lee} A.~T.~Lee {\it et al.},
Astrophys.\ J.\  {\bf 561} (2001) L1
[arXiv:astro-ph/0104459].

\bibitem{peloso} J.~Lesgourgues and M.~Peloso,
Phys.\ Rev.\ D {\bf 62} (2000) 081301
[arXiv:astro-ph/0004412].

\bibitem{lewisbridle}
 A.~Lewis and S.~Bridle,
''
arXiv:astro-ph/0205436.


\bibitem{Lisi}
E. Lisi, S. Sarkar, and F.L. Villante, {\it Phys. Rev.} {\bf D59} (1999)
123520.

\bibitem{mauskopf} P.~D.~Mauskopf {\it et al.}  [Boomerang Collaboration],
Astrophys.\ J.\  {\bf 536}, L59 (2000)
[arXiv:astro-ph/9911444].

\bibitem{McDonald}
J.McDonald, {\it Phys. Rev. Lett.} {\bf 84} (2000) 4798.

\bibitem{mcgaugh} S.~S.~McGaugh,
Astrophys.\ J.\  {\bf 541} (2000) L33
[arXiv:astro-ph/0008188].

\bibitem{melchiorri} A.~Melchiorri {\it et al.}  [Boomerang Collaboration],
Astrophys.\ J.\  {\bf 536} (2000) L63
[arXiv:astro-ph/9911445].

\bibitem{melk99} A.~Melchiorri, M.~V.~Sazhin, V.~V.~Shulga and N.~Vittorio,
Astrophys.\ J.\  {\bf 518} (1999) 562, [arXiv:astro-ph/9901220].

\bibitem{melksilk} A.~Melchiorri and J.~Silk,
arXiv:astro-ph/0203200.

\bibitem{miller} A.~D.~Miller {\it et al.},
Astrophys.\ J.\  {\bf 524}, L1 (1999)
[arXiv:astro-ph/9906421].

\bibitem{mortsell}
S.~Hannestad and E.~Mortsell,
Phys.\ Rev.\ D {\bf 66} (2002) 063508
[arXiv:astro-ph/0205096].

\bibitem{netterfield} C.~B.~Netterfield {\it et al.}  
[Boomerang Collaboration], arXiv:astro-ph/0104460.

\bibitem{carolina}
C.~J.~Odman, A.~Melchiorri, M.~P.~Hobson and A.~N.~Lasenby,
arXiv:astro-ph/0207286.

\bibitem{carolina2}
A. Melchiorri, C. J. Odman, astro-ph/0210606, PRD in press (2002).

\bibitem{orito}
M.~Orito, T.~Kajino, G.~J.~Mathews and R.~N.~Boyd,
arXiv:astro-ph/0005446.

\bibitem{pearson}
T. J. Pearson {\it et al.}, astro-ph/0205388, (2002).

\bibitem{PTW}S.\ Perlmutter, M.S. Turner, M. White, Phys.Rev.Lett. {\bf 83} 
670-673 (1999).                                                               

\bibitem{pryke} C.~Pryke, N.~W.~Halverson, E.~M.~Leitch, 
J.~Kovac, J.~E.~Carlstrom, W.~L.~Holzapfel and M.~Dragovan,
arXiv:astro-ph/0104490.

\bibitem{scott}
P. F. Scott {\it et al.}, astro-ph/0205380, (2002).

\bibitem{sz} U. Seljak, M. Zaldarriaga, 
M. 1996, \apj, 469, 437.


\bibitem{tegb97} M.~Tegmark,
Astrophys.\ J.\  {\bf 514}, L69 (1999)
[arXiv:astro-ph/9809201].

\bibitem{thx}M. Tegmark, A. J. S. Hamilton, Y. Xu, 
astro-ph/0111575 (2001)

\bibitem{torbet} E.~Torbet {\it et al.},
Astrophys.\ J.\  {\bf 521}, L79 (1999)
[arXiv:astro-ph/9905100].

\bibitem{wang} X. Wang, M. Tegmark, M. Zaldarriaga, astro-ph/0105091.

\bibitem{podariu} S. Podariu et al., Astrophys.J. 559 (2001) 9

\bibitem{cmiller} C. Miller et al., astro-ph/0112049 (2001).

\bibitem{jaffe} A. Jaffe et al., Phys.Rev.Lett. 86 (2001) 3475-3479.

\bibitem{ganga} Ganga K., Ratra B., Gundersen, J., Sugyiama, N. 1997,
ApJ, 484, 7.

\bibitem{peeblesratra} P.J.E. Peebles, B. Ratra, astro-ph/0207347.


\end{thebibliography}
\end{document}